\documentclass[fleqn,usenatbib]{mnras}

\usepackage {psfig}
\usepackage{amssymb}
\usepackage {graphicx}
\usepackage {epsfig}
\usepackage{mathrsfs}
\usepackage[english]{babel}
\usepackage{amsfonts}
\usepackage{fontenc}
\usepackage{textcomp}
\usepackage{microtype}
\usepackage[pdftex]{lscape}
\usepackage{amsmath}

\title[Discreteness effects and ROI]{Discreteness effects, $N-$body chaos and the onset of radial-orbit instability}
\author[Di Cintio and Casetti]{Pierfrancesco Di Cintio$^{1,2}$\thanks{E-mail:pierfrancesco.dicintio@fi.infn.it} and Lapo Casetti$^{1,2,3}$\\
$^{1}$Dipartimento di Fisica e Astronomia, Universit\`a di Firenze, via G.\ Sansone 1, I-50019 Sesto Fiorentino, Italy\\
$^{2}$INFN -  Sezione di Firenze, via G.\ Sansone 1, I-50019 Sesto Fiorentino, Italy\\
$^3$INAF - Osservatorio Astrofisico di Arcetri, largo Enrico Fermi 5, I-50125 Firenze, Italy
}
\hypersetup{draft}
\begin{document}
\date{Accepted...  Received...; in original form...}
\pubyear{0000}
\maketitle
\begin{abstract}
We study the stability of a family of spherical equilibrium models of self-gravitating systems, the so-called $\gamma-$models with Osipkov-Merritt velocity anisotropy, by means of $N-$body simulations. In particular, we analyze the effect of self-consistent $N-$body chaos on the onset of radial-orbit instability (ROI). We find that degree of chaoticity of the system associated to its largest Lyapunov exponent $\Lambda_{\rm max}$ has no appreciable relation with the stability of the model for fixed density profile and different values of radial velocity anisotropy. However, by studying the distribution of the Lyapunov exponents $\lambda_{\rm m}$ of the individual particles in the single-particle phase space, we find that more anisotropic systems have a larger fraction of orbits with larger $\lambda_{\rm m}$. \end{abstract}

\section{Introduction}
\label{intro}
Collisionless and spherical self-gravitating systems with equilibrium phase-space distribution functions $f$ associated to large degrees of radial anisotropy (i.e. the velocity distribution is dominated by orbits with low values of the angular momentum $J$) are prone to the so-called as radial-orbit instability, (hereafter ROI, e.g. see  \citealt{1992JETP...74..755P,2008gady.book.....B,2011TTSP...40..425M,2014dyga.book.....B}). The origin of this process, despite the the large efforts made from both the analytical (e.g. \citealt{1981SvA....25..533P,1985AJ.....90.1027M,1987MNRAS.224.1043P,1992JETP...74..755P,1990BAAS...22.1261S,1991MNRAS.248..494S,1992SvA....36..482P,1994LNP...433..143P,1994ApJ...434...94B,2006ApJ...637..717T,2011MNRAS.416.1836P,2015MNRAS.451..601P,2017MNRAS.470.2190P}) and the numerical sides (e.g. \citealt{1973A&A....24..229H,1985MNRAS.217..787M,1986ApJ...300..112B,1987IAUS..127..315M,1990MNRAS.242..576A,1990ApJ...354...33A,1997ApJ...490..136M,2002MNRAS.332..901N,2007MNRAS.375.1157B,2009ApJ...704..372B,2015MNRAS.447...97G}), it is still debated (for an extensive review see e.g. \citealt{2011TTSP...40..425M}), and a deep understanding of the ROI has not been reached.\\
\indent Usually, it is assumed that the degree of anisotropy of a given spherical model with density $\rho(r)$ is quantified by the so-called Fridman-Polyachenko-Shukhman stability indicator (\citealt{1981SvA....25..533P,1984sv...bookR....F}) defined by
\begin{equation}\label{index}
\xi\equiv\frac{2K_r}{K_t},
\end{equation}
where $K_r$ and $K_t=K_\theta+K_\phi$ are the radial and tangential components of the kinetic energy tensor that read
\begin{equation}\label{krkt}
K_r=2\pi\int\rho(r)\sigma^2_r(r)r^2{\rm d}r,\quad K_t=2\pi\int\rho(r)\sigma^2_t(r)r^2{\rm d}r.
\end{equation} 
In the expressions above, $\sigma^2_r$ and $\sigma^2_t$ are the radial and tangential components, of the velocity dispersion tensor, defined for a given phase-space distribution $f$ (see e.g. \citealt{2008gady.book.....B}) as  
\begin{eqnarray}\label{sigmas}
\sigma^2_{ij}(\mathbf{r})\equiv\frac{1}{\rho(\mathbf{r})}\int_V(v_i-\bar{v}_i)(v_j-\bar{v}_j)f(\mathbf{r},\mathbf{v}){\rm d}^3\mathbf{v}=\nonumber\\
=\overline{v_iv_j}-\bar{v}_i\bar{v}_j,
\end{eqnarray}
where $\rho(\mathbf{r})\equiv\int f(\mathbf{r},\mathbf{v}){\rm d}^3\mathbf{v}$ and the bars over the symbols indicate averaged quantities. $N-$body simulations of anisotropic systems seem to suggest that, albeit with some weak dependence on the specific equilibrium model and/or initial density profile, for $\xi>\xi_{s}\simeq 1.5\pm 0.2$ they are unstable and rapidly evolve from spherical towards flattened or triaxial shapes. However, analytical results based on the spectral analysis of perturbations obtained by \cite{1987MNRAS.224.1043P,1994ASSL..185.....P,1994LNP...433..143P,2019MNRAS.487..711R} seem to indicate instead that no critical value of the anisotropy indicator $\xi_s$, above which the model is unstable, exists but instead that, whenever the distribution function $f$ diverges for a value of the angular momentum $J\to 0$, there is always a spectrum of unstable modes with frequencies $\omega=0$ as an accumulation point, for non spherical perturbations of $\rho$. Moreover, independently of the actual relation between $\xi$ and the ROI, it is also unclear whether the onset of the instability is a collective effect (and therefore connected to the {\it global} phase-space properties of the system), or a local effect (e.g. connected to its granularity) ``amplified" due to the long-range nature of the $1/r^2$ Newtonian force.\\
\indent In a series of papers by \cite{2011MNRAS.414.3298N,PLA:9846257,2017MNRAS.468.2222D} the ROI was investigated in Modified Newtonian Dynamics (MOND, \citealt{1983ApJ...270..365M,1984ApJ...286....7B}) and additive long-range forces of the form $1/r^\alpha$ with $-1\leq\alpha<3$. What was found in the context of MOND is that, on one hand, MOND systems are always more likely to undergo ROI than their equivalent Newtonian Systems (ENS, i.e. Newtonian systems where the baryonic component has the same phase-space distribution as the parent MOND model). On the other hand, MOND systems are able to support a larger amount of kinetic energy stored in radial-orbits than single-component Newtonian systems with the same density distribution and without a spherical  Dark Matter halo, whose presence in Newtonian gravity has usually a mild stabilizing effect against ROI, as found in numerical simulations (\citealt{1991ApJ...382..466S,1997ApJ...490..136M,2002MNRAS.332..901N}).\\
\indent As to $1/r^\alpha$ forces, it emerged that, independently of the specific value of $\alpha$, isotropic models are always associated with monotonic phase-space distribution functions $f$, while all models with significantly non-monotonic $f$ are violently unstable. Numerical simulations showed that in general, for fixed density $\rho$, systems with lower values of $\alpha$ are able to support larger amounts of radial anisotropy (i.e., higher values of $\xi$), unstable models with low values of $\alpha$ have more triaxial end-products, while models with larger values of $\alpha$ (i.e. for which the inter-particle force is ``more local"), even when critically unstable, tend to remain closer to spherical. All these results suggest that the ROI is a feature of systems interacting with long-range forces (either additive, like the $1/r^\alpha$ studied by \citealt{2011IJBC...21.2279D,2013MNRAS.431.3177D,PLA:9846257,2017MNRAS.468.2222D}, or associated to non-linear field equations such as MOND), and not restricted to the Newtonian force only.\\
\indent Other important points are how much ROI is connected to the chaoticity of the gravitational $N-$body problem with different initial conditions in velocity space (i.e. different choices of anisotropic $f$), and whether the presence of externally and/or self-consistently induced ``noise" and dissipation along individual particle orbits affect the onset of the instability. Muzzio and collaborators (\citealt{1995ASIB..336..537C,1995ApJ...440....5C,1996ApJ...456..274C,2012MNRAS.423.1955Z,2014MNRAS.438.2871C}) studying orbits in self-consistent simulations of cold collapses and smooth potential spherical systems with non-radial perturbations found that, in general, isotropic velocity distributions suppress chaoticity (i.e. on average, orbits have smaller Lyapunov exponents), while strongly anisotropic initial conditions are always associated to a larger fraction of chaotic orbits.\\
\indent For what concerns the role of dissipation, \cite{2010MNRAS.405.2785M} suggested that an effective dissipative mechanism acting on the single orbit is a necessary condition for ROI to happen, even for purely radial models where orbits have only one degree of freedom and can neither precede nor librate (as required for example in the original interpretation of ROI of \citealt{1994LNP...433..143P}). The source of effective energy dissipation can be traced back to  the discreteness of the system as well as to non-gravitational physics, if present, cosmological factors or the effects of the Dark Matter distribution. In numerical simulations one has an extra source of effective dissipation in numerical errors. \\
\indent In this work we explore the relation between ROI and chaos by means of $N-$body simulations. We follow the evolution of a family of $\gamma-$models with different degrees of initial radial anisotropy $\xi_0$, analyzing their degree of chaoticity quantified by their largest Lyapunov exponent $\Lambda_{\rm max}$ and studying their spatial properties.\\ 
\indent The paper is structured as follows. In Section 2 we introduce the self-gravitating system models and the set-up of the initial conditions for the simulation, and we introduce the indicators that quantify the dynamical stability of numerical models. In Section 3 present the results of our numerical simulations on the evolution of self-consistent systems characterized by different degrees of radial anisotropy, and study the structural properties of their final states as functions of $\xi_0$, $N$, ecc. The main results are finally discussed and summarized in Section 4. 
\section{Setting the stage}\label{sec:1}
\subsection{Models}
We consider the the so-called $\gamma-$Model family of spherical density profiles, introduced by \cite{1993MNRAS.265..250D} (see also \citealt{1994AJ....107..634T}), given by
\begin{equation}\label{dehnen}
\rho(r)=\frac{3-\gamma}{4\pi}\frac{Mr_c}{r^\gamma(r+r_c)^{4-\gamma}},
\end{equation}
where $M$ is the total mass, $r_c$ the core radius, and $0\leq\gamma<3$ the so-called logarithmic density slope\footnote{Note that for $\gamma=2$ and $\gamma=1$ one recovers the  \cite{1983MNRAS.202..995J} and \cite{1990ApJ...356..359H} models, respectively.}. With such a choice, we can model systems ranging from those characterized by a flat core ($\gamma=0$) up to those with a strong central cusp ($\gamma\to 3$).\\
\indent In order to generate the initial conditions for the $N-$body simulation, we first of all generate the particle position by sampling the cumulative mass function associated to the density (\ref{dehnen})
\begin{equation}
M(r)=M\left (\frac{r}{r+r_c}\right )^{3-\gamma}
\end{equation}
in the standard way. Once the positions are obtained, the velocities are assigned with a rejection method from the anisotropic phase-space distribution function $f(Q)$ with Osipkov-Merritt radial anisotropy (hereafter OM, \citealt{1979SvAL....5...42O,1985AJ.....90.1027M}, see also \citealt{1996ApJ...471...68C,1999ApJ...520..574C}), given by the reparametrization of the standard Eddington inversion formula for isotropic systems \cite{1916MNRAS..76..572E} as
\begin{equation}\label{OM}
f(Q)=\frac{1}{\sqrt{8}\pi^2}\int_Q^{0}\frac{{\rm d}^2\rho_a}{{\rm d}\Phi^2}\frac{{\rm d}\Phi}{\sqrt{\Phi-Q}},
\end{equation}
where $\Phi$ is the gravitational potential, that for the density profile (\ref{dehnen})
reads
\begin{equation}
\Phi(r)=-\frac{GM}{(2-\gamma)r_c}\left[1-\left(\frac{r}{r+r_c} \right)^{2-\gamma} \right]\quad {\rm for}\quad \gamma\neq 2
\end{equation}
and 
\begin{figure*}
  \includegraphics[width=0.95\textwidth]{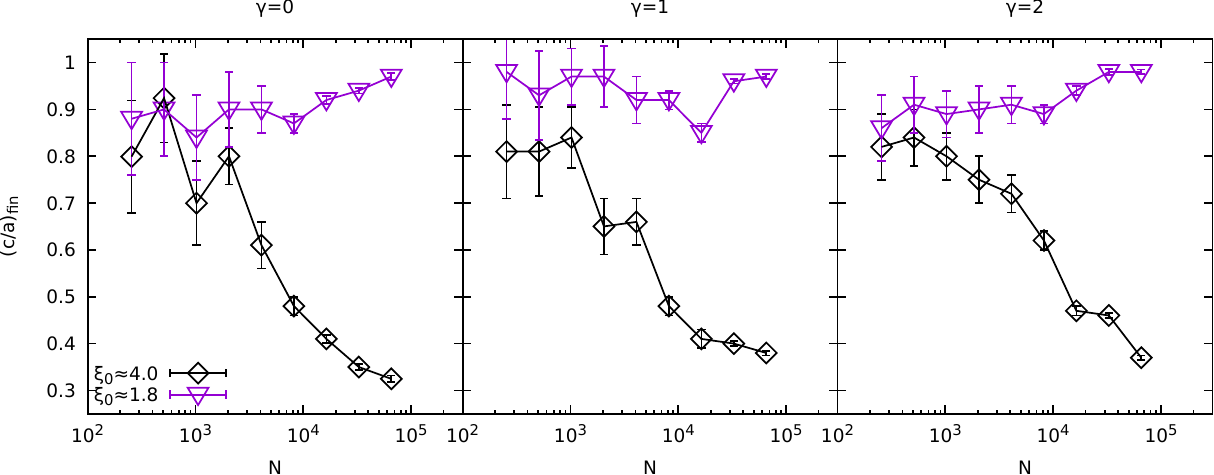}
\caption{Axial ratio $c/a$ at $t/t_*=200$ for radially anisotropic Dehnen models with (from left to right) $\gamma=0$, 1 and 2, as function of the particle number $N$, for $\xi_0\approx 4$ (diamonds) and $\xi_0\approx 1.8$ (downward triangles).}
\label{figxi}       
\end{figure*}
\begin{equation}
\Phi(r)=\frac{GM}{r_c}\ln\frac{r}{r+r_c}\quad {\rm for}\quad \gamma= 2.
\end{equation}
In the expressions above $Q=E+{J^2}/{2r_a^2}$, and $E$ and $J$ are the particle's energy and angular momentum per unit mass, respectively, $r_a$ is the anisotropy radius, and $\rho_a$ the augmented density, defined by 
\begin{equation}\label{augmented}
\rho_a(r)\equiv\left(1+\frac{r^2}{r_a^2}\right)\rho(r).
\end{equation}
The anisotropy radius $r_a$ controls the extent of anisotropy of the model, that is, the velocity-dispersion tensor is nearly isotropic inside $r_a$, and more and more radially anisotropic for increasing $r$. Therefore, small values of $r_a$ correspond to more radially anisotropic systems, and thus to larger values of the anisotropy parameter $\xi$.\\
\indent Note that, for OM models, the velocity dispersions can also be evaluated (see e.g. \citealt{1995MNRAS.276.1131C}) from
\begin{eqnarray}\label{sigmas2}
\rho(r)\sigma_r^2(r)&=&-\frac{r_a^2}{r_a^2+r^2}\int_r^\infty\rho_a(r)\frac{{\rm d}\Phi}{{\rm d}r}{\rm d}r;\nonumber\\
\rho(r)\sigma_t^2(r)&=&\frac{2r_a^2}{r_a^2+r^2}\rho(r)\sigma_r(r)^2,
\end{eqnarray}
instead of computing the radial and tangential components of the tensor defined in Equations (\ref{sigmas}). In the numerical realizations of the OM models used as initial conditions the value of the initial anisotropy parameter $\xi_0$ is quantified by solving numerically Equations (\ref{sigmas2}) for $\sigma_r$ and $\sigma_t$ and then Equations (\ref{krkt}), and also, in discrete form, by simple particle counts as
\begin{equation}
\sigma_r^2(R)=\frac{1}{N_R}\sum_{i=1}^{N_R}(v_{r,i}-\bar{v_r})^2;\quad \sigma_t^2(R)=\frac{1}{N_R}\sum_{i=1}^{N_R}(v_{t,i}-\bar{v_t})^2,
\end{equation}
where $\bar{v_r}$ and $\bar{v_t}$ are the radial and tangential mean velocities in the radial bin $R+dr$, respectively, and $N_R$ is the number of particles there contained.\\
\indent Throughout this paper we always refer to the values of $\xi_0$ obtained evaluating $K_r$ and $K_t$ with particle counts, as they do not differ, for $N>500$, for more than the 5\% (largest deviance for the smallest models considered with $N=256$) from the values obtained with the (semi-)analytical procedures. We verified that uncertainty on the values of $\xi_0$ obtained by direct sum from the $N-$body realization depends mainly on $N$ (data not shown), while it has little to none dependence on the specific form of $f$.
\subsection{Numerical methods}
In our $N-$body simulations we solve the particles equations of motion 
\begin{equation}\label{eom}
\ddot{{\mathbf r}}_i=-Gm\sum_{j=1}^N\frac{{\mathbf r}_i-{\mathbf r}_j}{||{\mathbf r}_i-{\mathbf r}_j||^3}.
\end{equation} 
with the symplectic integrator with adaptive order of \cite{1991CeMDA..50...59K} with fixed time-step $\Delta t$.\\ 
\indent Since we have to integrate models with different density distributions, associated to different choices of $\gamma$ and characterized in principle by different crossing time scales for the same value of the total mass $M$, we use a common normalization of particle positions and velocities. In the simulations presented in this work, all positions are in units of the initial half mass radius (i.e. the radius containing $M/2$ at $t=0$), that for a $\gamma-$models reads
\begin{equation}
r_*=\frac{r_c}{2^{1/(3-\gamma)}-1}.
\end{equation}
The dynamical time and velocity scales are then fixed as
\begin{equation}
t_*=\sqrt{2r_*^3/GM};\quad v_*=r_*/t_*,
\end{equation}
so that the gravitational constant and individual particle masses are set to $G=1$ and $m=2/N$. In these units we adopt a fixed $\Delta t=5\times 10^{-3}$ and an optimal (see e.g. \citealt{2005ARep...49..470R,2011EPJP..126...55D}) softening parameter\footnote{The softening length $\epsilon_{\rm soft}$ is such that the potential at distance $r$ from a particle of mass $m$ is $\phi(r)=-Gm/\sqrt{r^2+\epsilon_{\rm soft}^2}$.} $\epsilon_{\rm soft}=5\times 10^{-3}$ and we use a 3rd order integration scheme. We settled to such combination of $\epsilon_{\rm soft}$ and $\Delta t$ as a 
\begin{figure*}
  \includegraphics[width=\textwidth]{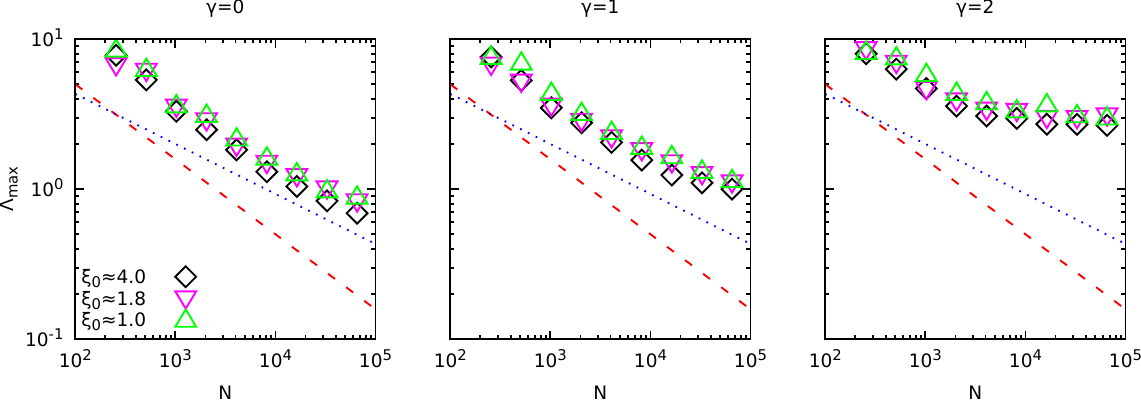}
\caption{Maximal Lyapunov exponent $\Lambda_{\rm max}$ as function of the number of particles $N$ for $\gamma-$models with (from left to right) $\gamma=0$, 1 and 2, and different values initial degrees of anisotropy $\xi_0\approx 4,$ 1.8 and 1. The dotted and dashed lines mark the $N^{-1/3}$ and $N^{-1/2}$ trends, respectively.}
\label{figlmax}       
\end{figure*}
further reduction of the softening parameter (and the associated timestep) will not result in significant changes in the evolution of the axial ratios and the estimates of Lyapunov exponents for a given initial condition.\\
\indent In order to evaluate the Lyapunov exponents we also solve the variational equations associated to the dynamics given by Eq. (\ref{eom}) for the tangent vectors $\mathbf{w}_i$ (\citealt{1971JCoPh...8..449M,1993ApJ...415..715G,2002ApJ...580..606H,2016MNRAS.459.2275R,2019arXiv190108981D})
\begin{equation}\label{var}
\ddot{{\mathbf w}}_i=-Gm\sum_{j=1}^N\left[\frac{{\mathbf w}_i-{\mathbf w}_j}{||{\mathbf r}_i-{\mathbf r}_j||^3}-3({\mathbf r}_i-{\mathbf r}_j)\frac{({\mathbf w}_i-{\mathbf w}_j)\cdot({\mathbf r}_i-{\mathbf r}_j)}{||{\mathbf r}_i-{\mathbf r}_j||^5}\right].
\end{equation}
We evaluate numerically an estimate of the (finite time) largest Lyapunov exponent of an $N-$body model with the standard  Benettin-Galgani-Strelcyn method (\citealt{1976PhRvA..14.2338B}, see also \citealt{2002ocda.book.....C,2007PhRvL..99m0601G,2013JPhA...46y4005G}) as 
\begin{equation}\label{lmax}
\Lambda_{\rm max}(t) =\frac{1}{L\Delta t}\sum_{k=1}^L\ln\frac{W(k\Delta t)}{W_0}~,
\end{equation}
for a (large) time $t=L\Delta t$, where $W$ is the norm of the $6N$-dimensional vector 
\begin{equation}\label{w6n}
\mathbf{W}_{6N}=(\mathbf{w}_i,\dot{\mathbf{w}}_i,...\mathbf{w}_N,\dot{\mathbf{w}}_N),
\end{equation}
where $W_0$ is the value of such norm at $t=0$. Following \cite{1976PhRvA..14.2338B}, in order to improve the convergence, the vector $\mathbf{W}_{6N,6}$ is periodically renormalized to $W_0$. In all simulations presented here, the renormalization procedure is done every $10\Delta t$. However, The value attained at time $t$ by $\Lambda_{\rm max}$ is independent of the frequency of this operation and the value of $W_0$, that we fix to unity in all simulations shown here. In addition, we have also computed in some runs the largest Lyapunov exponents $\lambda_{m, i}$ of {\it individual} particles by evaluating expressions (\ref{lmax}) and (\ref{w6n}) in the 6-dimensional phases-spaces of each particle moving in the (time-dependent) potential of all the others. Note that, in this latter case, in the renormalization procedure each tangent vector $\mathbf{w}_i$ is renormalized to its initial size $w_{0,i}$.  
\section{Simulations and results}\label{sec:2}
\cite{1996ApJ...456..274C} interpreted the ROI as a mechanism that transforms loop orbits into box orbits (typical constituents of triaxial systems) arising in anisotropic spherical models subjected to small non-spherical perturbations, followed by a ``transition'' from quasi-regular to chaotic motion. This has led to speculate that in general, models with larger values of the anisotropy parameter $\xi$ may be somewhat associated to larger degrees of chaos.\\
\indent In order to explore this matter further, in a first set of numerical experiments we have evaluated $\Lambda_{\rm max}$ for different choices of the density profile, of the initial anisotropy (quantified by $r_a$ or $\xi_0$) and of the number of particles $N$. All numerical simulations have been extended up to $t/t_*=200$, that on average is larger than the typical timescale on which the ROI sets in (see \citealt{1997ApJ...490..136M,2002MNRAS.332..901N,2011MNRAS.414.3298N,2017MNRAS.468.2222D}) that is usually around $10t_*$. During the numerical integrations we have computed also the evolution of $\xi$ and of the minimum to maximum and intermediate to maximum axial ratios $c/a$ and $b/a$. The latter have been evaluated in the standard way (see e.g.\ \citealt{1997ApJ...490..136M,2013MNRAS.431.3177D,2017MNRAS.468.2222D}) by computing at the rank two tensor 
\begin{equation}\label{inertia}
I_{ij}\equiv \sum_{k} m_k r_i^{(k)}r_j^{(k)},
\end{equation}
related to the inertia tensor of the system by ${\rm Tr}(I_{ij})\delta_{ij}-I_{ij}$.
The sum in (\ref{inertia}) has been limited to the particles inside the sphere of Lagrangian radii $r_{90}$, $r_{70}$ and $r_{50}$, (i.e, the radius of the sphere containing 90\%, 70\%  and 50\% of the total mass of the system, respectively). The matrix $I_{ij}$ is iteratively diagonalized with the standard {\sc LAPACK} routines, with tolerance set to 0.1\%, in order to obtain its eigenvalues $I_{11}\geq I_{22}\geq I_{33}$. For a heterogeneous density distribution stratified over concentric and coaxial ellipsoidal surfaces of semi-axes $a\geq b\geq c$, we would obtain $I_{11}=Aa^2$, $I_{22}=Ab^2$ and $I_{33}=Ac^2$, where $A$ is a constant depending on the density profile. Once the three $I_{ii}$ are computed the fiducial axial ratios are obtained as $b/a=\sqrt{I_{22}/I_{11}}$ and $c/a=\sqrt{I_{33}/I_{11}}$, 
\begin{figure}
  \includegraphics[width=\columnwidth]{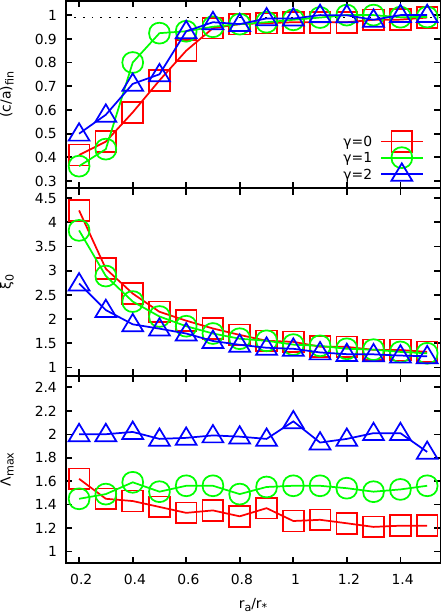}
\caption{Final minimum to maximum semi-axis ratio $(c/a)_{\rm fin}$ at $t/t_*=200$ (top panel), initial anisotropy parameter $\xi_0$, and maximal Lyapunov exponent $\Lambda_{\rm max}$ (bottom panel) for three families of radially anisotropic Hernquist models with $\gamma=0$, 1 and 2, as a function of the initial anisotropy radius $r_a$, for $N=20000$. The dotted line in the upper panel marks the bona-fide axial ratio 0.99 used as a threshold for stability of the model.}
\label{figca1}       
\end{figure}
so that the ellipticities in the principal planes are $\epsilon_1=1-\sqrt{I_{22}/I_{11}}$ and $\epsilon_2=1-\sqrt{I_{33}/I_{11}}$. The procedure is carried out for a number of snapshots of the numerical simulation (typically one every $5\Delta t$), and the final value of the minimum axial ratio $(c/a)_{\rm fin}$ is obtained by averaging over the values attained by $c/a$ over the last $40t_*$, with error corresponding to the standard deviation of the averaging operation.\\
\indent In Figure \ref{figxi} we show the final minimum-to-maximum axis ratio $(c/a)_{\rm fin}$ as a function of the number of particles $N$ for strongly anisotropic ($\xi_0\approx 4$) and mildly anisotropic ($\xi_0\approx 1.8$) unstable models with, from left to right, $\gamma=0,$ 1 and 2. Clearly, for the more anisotropic model the final value of $(c/a)$ depends strongly on the number of particles $N$, while in the less anisotropic cases $c/a$ attains similar values over a three decades span in $N$. In all cases however, the error bars decrease with increasing $N$ in the same way, as the both averaging and the diagonalization procedures have the same dependence on $N$. Remarkably, the values of $(c/a)_{\rm fin}$ are larger than $\approx 0.3$ (corresponding to an E7 system) in agreement with previous numerical studies on cold collapses and ROI (see e.g. \citealt{2002MNRAS.332..901N,2006MNRAS.370..681N,2011MNRAS.414.3298N,2013MNRAS.431.3177D,2017MNRAS.468.2222D} and references therein)\\
\begin{figure}
  \includegraphics[width=\columnwidth]{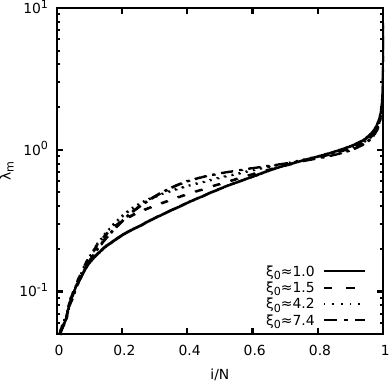}
\caption{Ordered plot of the single particle largest Lyapunov exponents $\lambda_{\rm m}$ for different $\gamma=1$ models with $N=20000$ and initial value of the orbital anisotropy $\xi_0\approx 7.4,$ 4.2, 1.5 and 1.}
\label{figord}       
\end{figure}
\indent The largest Lyapunov exponent $\Lambda_{\rm max}$ for these models, as well as for the associated isotropic systems, are shown in Figure \ref{figlmax}. Remarkably, little dependence on $\xi_0$ is found at fixed $N$: $\Lambda_{\rm max}$ has the same scaling with the number of particles $N$ for every value of $\gamma$. The typical error on the values of $\Lambda_{\rm max}$, obtained as the width of the oscillations of the time series in Eq. (\ref{lmax}) between for $180t_*\leq t\leq 200t_*$, steadily decreases with $N$ ranging from 0.5 to 0.01 for $N=10^2$ to $N=10^5$ (error bars are not included as they are roughly the size of the symbols or smaller). We find that the trend of the maximal Lyapunov exponent with the system size is compatible with the expected power law decay (see e.g. \citealt{1986A&A...160..203G,2009A&A...505..625G,2013AN....334..800O}) $\Lambda_{\rm max}\propto N^{-\alpha}$ with $\alpha$ between 1/2 and 1/3, (at least for $\gamma\lesssim 1.8$, see also similar plots in \citealt{2019arXiv190108981D,modest19}), while at larger values of the logarithmic density slope a saturation at large $N$ appears (see the $\gamma=2$ case in Fig. \ref{figlmax}), that is, models with steeper central density cusps have in general a larger degree of chaos  regardless of the amount of radial anisotropy.\\
\indent In Fig. \ref{figca1} we present the values attained at $t=200$ by the ratio of minimum to maximum semiaxes $c/a$ within $r_{90}$ (top panel), the initial anisotropy parameter $\xi_0$ (middle panel) and the largest Lyapunov exponent $\Lambda_{\rm max}$ (bottom panel) as a function of the initial anisotropy radius $r_a$, for $N=20000$ and the same three values of $\gamma=0,$ 1 and 2. Consistently with previous numerical results (e.g. see \citealt{1985MNRAS.217..787M,1997ApJ...490..136M,2002MNRAS.332..901N,2009ApJ...704..372B,2011MNRAS.414.3298N,2017MNRAS.468.2222D}), we observe that sensible deviations from the spherical symmetry appear for the 
\begin{figure*}
  \includegraphics[width=\textwidth]{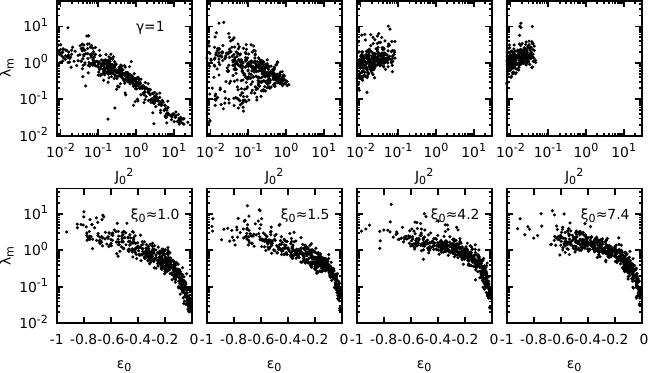}
\caption{Single particle largest Lyapunov exponents $\lambda_{\rm m}$ for the same models of Fig. \ref{figord} as function of their initial specific angular momentum $J^2_0$ (top  row) and energy $\mathcal{E}_0$ (bottom row).}
\label{figscatter}       
\end{figure*}
cases with $\xi_0\gtrsim 1.5$, corresponding approximately to $r_a/r_*\approx 1.1$ for $\gamma=0$, 0.9 for $\gamma=1$ and 0.8 for $\gamma=2$, (see e.g. \citealt{1996ApJ...471...68C}).\\
\indent Remarkably, we note that the value of the finite time largest Lyapunov exponent $\Lambda_{\rm max}$ is almost independent on the amount of radial anisotropy (and the specific values of $r_a$) of the models, settling with narrow variation range at $\Lambda_{\rm max}\approx 1.4\pm 0.2$ for $\gamma=0$, $1.5\pm 0.15$ for $\gamma=1$, and $2\pm 0.1$ for $\gamma=2$, with typical error bars of the order of 0.015 for $N=20000$.\\
\indent We observe that only in the case of $\gamma = 0$, that is, a cored density profile, a slight decrease of $\Lambda_{\rm max}$ with $r_a$ is detected, meaning that more radially anisotropic systems are, in this case, slightly more chaotic; the effect is however small. We have also checked whether this behaviour is stable for different choices of the softening length $\epsilon_{\rm soft}$, without finding any significant indication that for increasing or decreasing values of $\epsilon_{\rm soft}$, $\Lambda_{\rm max}$ develops a trend with $\xi_0$. However, in general, using smaller values of the softening length at fixed $\Delta t$ yields larger values of $\Lambda_{\rm max}$. In addition, we have performed runs for decreasing softening length {\it and} timesteps, so that $\epsilon_{\rm soft}$ remains smaller than the mean interparticle distance. What was found is that for large $N$ the trend of $\Lambda_{\rm max}$ with $N$ remains unchanged, while the slope becomes somewhat flatter at small values of $N$. In general, for $\epsilon_{\rm soft}\leq 3\times 10^{-4}$ in units of $r_c$, no appreciable change in $\Lambda_{\rm max}$ for appropriately small $\Delta t$ could be observed (see also \citealt{2019arXiv190108981D}). For a more detailed discussion of how the softening of the gravitational force influences the values attained by the finite time Lyapunov exponents see \cite{1993ApJ...415..715G,2002MNRAS.331...23E,2019MNRAS.484.1456E}.\\
\indent  We made some further test runs with larger and smaller numbers of particles finding that, independently on $\gamma$, the constant trend of $\Lambda_{\rm max}$ with $r_a$ (or $\xi_0$) persists at larger systems sizes while, in general, one finds systematically smaller values of $\Lambda_{\rm max}$ for larger $\xi_0$ when $N\lesssim 5000$ (data not shown).\\
\indent The fact that the value attained by the finite time largest Lyapunov exponents is seemingly unrelated to the initial amount of orbital anisotropy of initial conditions with the {\it same} density distribution and number of particles $N$ implies that also the associated Lyapunov time $\tau_L\propto\Lambda_{\rm max}^{-1}$ is scarcely, if not at all, influenced by $\xi_0$. However, this does not rule out, in principle, that that the full Lyapunov spectra of models with different initial anisotropy $\xi_0$ may differ substantially though having comparable maxima, being therefore associated to different distributions of instability time-scales connected to the inverse of the Lyapunov exponents $\Lambda_i$.\\
\indent Computing the full Lyapunov spectrum for a self-consistent $N-$body system with large $N$ is extremely expensive in terms of memory and computational time, as it involves $\mathcal{O}(N^3)$ operations, so it has been attempted successfully only for one dimensional toy models (\citealt{2019JPhA...52A4001D}). In order to get insight on more detailed properties of the chaotic dynamics of these systems we computed, in some selected runs, the largest Lyapunov exponent $\lambda_{i,{\rm m}}$ in the six-dimensional phase-space of each simulation particle in the (time dependent) potential of the other $N-1$ particles, and we then extracted their cumulative distribution (i.e., the ordered plot of $\lambda_{i,{\rm m}}$). 
\begin{figure*}
  \includegraphics[width=0.95\textwidth]{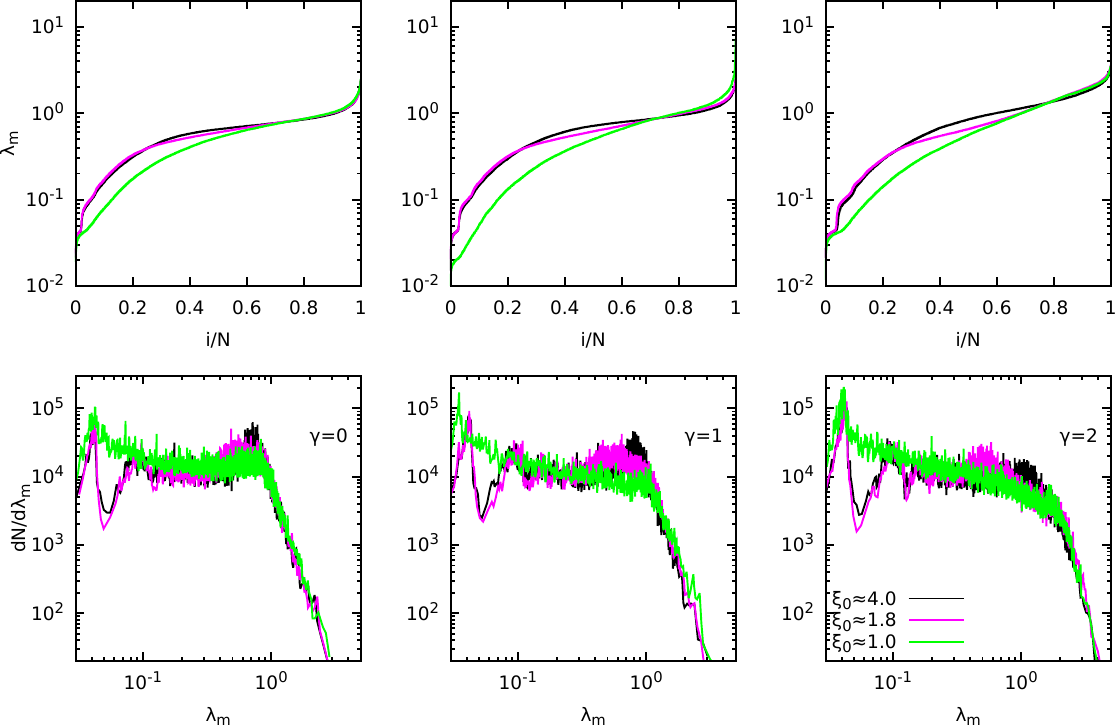}
\caption{For the same values of $\xi_0$ and $\gamma$ of the models in Fig. \ref{figlmax}: cumulative distributions of single particle largest Lyapunov exponents (top row), and their differential distribution ${\rm d}N/{\rm d}\lambda_{\rm m}$ (bottom row) for the $N=16384$ case.}
\label{figdist}       
\end{figure*}
Although the latter quantity is not the Lyapunov spectrum of the $N$-body system, it still gives some detailed information on the distribution of instability times of the different single-particle orbits. \\
\indent In Fig.\ \ref{figord} we show the cumulative distributions of the single particle Lyapunov exponents for all  particles in $\gamma=1$ (Hernquist) models with initial values of the anisotropy between 7.4 (highly unstable, close to consistency limit) and 1 (stable isotropic model). In all cases the curves peak\footnote{Note that the largest Lyapunov exponent of the parent full $N-$body system $\Lambda_{\rm max}$ has a significantly different value (in this case $\Lambda_{\rm max}\approx 1.5$), as the two quantities have, in principle, a different meaning.} at $\lambda_m\approx 11$, but have remarkably different slopes. In particular, the cases with larger amounts of initial anisotropy (i.e. $\xi_0\approx 7.4$ and 4.2, dotted-dashed and dashed lines) have systematically larger values than the isotropic model (solid line) over a fraction of roughly the 80\% of the sampled single particle orbits. In general, the values of the single particle Lyapunov exponents $\lambda_{\rm m}$ have the same dependence on the initial orbit energy (per unit mass) $\mathcal{E}_0$ (see the scatter plots in Fig. \ref{figscatter}, lower panel) for different initial choices of $\xi$ at fixed $N$ and $\gamma$, with initially less bound particles (i.e. $\mathcal{E}_0\to 0$) associated to smaller exponents.
 Vice versa, due to the intrinsically different distribution of angular momentum $J_0$ 
in OM models with different $\xi_0$, the dependence of $\lambda_{\rm m}$ on $J_0$ varies strongly with $\xi_0$ with a remarkable change in decreasing with increasing trend of $\lambda_{\rm m}$ with $J_0^2$ (see upper panels, same figure).\\
\begin{figure}
  \includegraphics[width=\columnwidth]{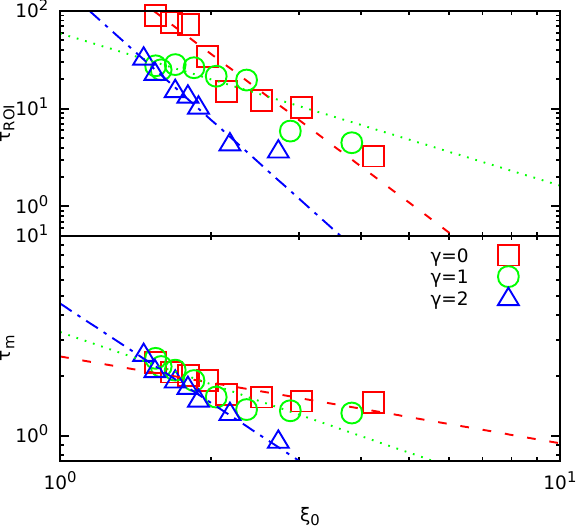}
\caption{Time of ignition of the ROI $\tau_{\rm ROI}$ (top panel) and collective particle Lyapunov time $\tau_m$ (bottom) as function of the initial stability indicator $\xi_0$, for model with $\gamma=0$ (squares), 1 (circles) and 2 (triangles). The dashed lines are the best-fit curves.}
\label{figtime}       
\end{figure}
\indent In addition to the cumulative distributions we also evaluate its derivative ${\rm d}N/{\rm d}\lambda_{\rm m}$ (i.e. the differential distribution of the single particle largest Lyapunov exponents) for different values of $N,$ $\gamma$ and $\xi_0$. In Figure \ref{figdist} we show the cumulative distributions (upper panels) and their associated differential distribution (lower panels) for different choices of $\xi_0$ for  models with $N=16384$ and $\gamma=0,$ 1 and 2. For all logarithmic density slopes $\gamma$ the cumulative distribution of the isotropic case (green/light gray curve) systematically remains below the corresponding curve for the anisotropic models (cfr. also Fig. \ref{figord}) for roughly the 75\% of the total number of particles, while, again, the values of the maximum $\lambda_{\rm m}$ do not differ significantly. The differential distributions of single particle largest Lyapunov exponents are, independently of $\gamma$, peaked at low $\lambda_{\rm m}$ for isotropic systems, while present a second peak at around $\lambda_{\rm m}\approx 1$ for the anisotropic cases. The position of the second peak moves towards larger values of $\lambda_{\rm m}$ for increasing $\xi_0$. Such behaviour of the distribution of Lyapunov exponents implies that the distribution of single particle Lyapunov times defined as $\tau_l\equiv\lambda_{\rm m}^{-1}$ peaks at {\it smaller} times for increasing anisotropy. We then define as collective Lyapunov time $\tau_{\rm m}$ the inverse of the value at which ${\rm d}N/{\rm d}\lambda_{\rm m}$ has its relative maximum, that is, the time scale over which the largest fraction of orbits can develop instabilities. It is then interesting to establish how does $\tau_{\rm m}$ scale with $\xi_0$ and what is its relation to the ROI time scale $\tau_{\rm ROI}$ (i.e., the time at which $c/a$ starts to depart significantly from unity). In Figure \ref{figtime} we show for the same choices of $\gamma$ and $N=20000$ the dependence of $\tau_{\rm ROI}$ and $\tau_{\rm m}$ with $\xi_0$ (symbols). We find that both characteristic time scales are well fitted by a power law (dashed lines) with (as expected) lower values of both times for larger values of $\xi_0$. In particular, we observe that at fixed $\xi_0$ $\tau_{\rm ROI}$ is systematically larger than $\tau_{\rm m}$.        
\section{Discussion and conclusions}\label{sec:3}
In this paper we have continued our study on the effect of $N-$body chaos and discreteness ``noise'' on the evolution of orbits and instabilities in spherical self-gravitating systems, following \cite{2019arXiv190108981D}. We have investigated the onset of the radial orbit instability in a family of Osipkov-Merritt-Dehnen models for various values of the Friedman-Polyachenko-Shukhman index $\xi$ and different sizes $N$ and logarithmic central density slopes $\gamma$. In particular, we have studied the trend of the largest Lyapunov exponent $\Lambda_{\rm max}$ with $N$ and $\xi_0$. We find that $\Lambda_{\rm max}$ has little to no dependence on $\xi_0$ at fixed system size $N$. This suggests that the ROI (and its typical time-scale) has no relation to the degree of {\it collective} chaoticity of the model. Vice versa, studying the single particle Lyapunov exponents in each particle's six-dimensional phase-space reveals that more anisotropic systems have in general (at fixed $\gamma$ or $N$) a larger fraction of orbits with large values of their largest Lyapunov exponents $\lambda_{\rm m}$. We interpret this as a more ``local'' (i.e., related to the evolution of orbits) rather than ``collective'' origin for the ROI. Moreover, we observe that both the time scale at which the instability sets in and the typical single particle Lyapunov time (i.e., the reciprocal of $\lambda_{\rm m}$ associated to the peak of the distribution of Lyapunov exponents) scale as a power law of the initial amount of anisotropy $\xi_0$ at fixed $N$ and for all the explored values of $\gamma$. However, such times differ of about a factor of ten, so that it is not obvious that they are somewhat related and the question remains open.     
\section*{Acknowledgements}
This work is part of MIUR-PRIN2017 project \textit{Coarse-grained
description for non-equilibrium systems and transport phenomena
(CO-NEST)} n.201798CZL whose partial financial support is acknowledged. The anonymous Referee is warmly acknowledged for his/her comments that helped improving the presentation of our results.
\bibliographystyle{mnras}
\bibliography{biblio.bib}
\end{document}